\documentstyle[12pt]{ioplppt}
\eqnobysec
\begin{document}
\jl{1}
\title{On-line learning of non-monotonic rules 
by simple perceptron}[On-line learning of non-monotonic rules]
\author{Jun-ichi Inoue\dag\ftnote{3}
{E-mail address: jinoue@stat.phys.titech.ac.jp},
Hidetoshi Nishimori\dag \,\,and 
Yoshiyuki Kabashima\ddag}
\address{\dag\ Department of Physics,
Tokyo Institute of Technology,
Oh-okayama, Meguro-ku,
Tokyo 152,
Japan
}
\address{\ddag\ Department of Computational Intelligence and Systems Science,
Interdisciplinary Graduate School of Science and Engineering,
Tokyo Institute of Technology, Yokohama 226, Japan}
\begin{abstract}
We study the generalization ability of a simple perceptron
which learns unlearnable rules.   The rules are presented by
a teacher perceptron with a non-monotonic transfer function.
The student is trained in the on-line mode. 
The asymptotic
behaviour of the generalization error is estimated 
under various conditions. 
Several learning strategies 
are proposed and improved to obtain 
the theoretical lower bound of 
the generalization error. 
\end{abstract}
\maketitle
\makeatletter
\renewcommand{\theequation}{%
\thesection.\arabic{equation}}
\@addtoreset{equation}{section}
\makeatother
\section{Introduction}
One of the important features of feed-forward neural networks is
their ability of learning a rule from examples
\cite{Hertz,Watkin,Opper}.
The student network can adopt its synaptic weights
following a set of examples given from the teacher network
so that
it can make predictions on the output for an input
which has been not shown before.  
Learning of unlearnable rules by a perceptron
is a particularly interesting issue
because the student usually does
not know the structure of the teacher in the real world.
For machine learning, it is important to
improve the learning scheme
and minimize the prediction error even if it is impossible
to exactly reproduce the input-output relation of the teacher.
Only a few papers have appeared concerning
learning  of unlearnable rules
where the teacher and the student have different
structures \cite{Kim,Saad95,Watkin92}.

In this paper we study the generalization ability
of a simple perceptron using the on-line algorithm
from a teacher perceptron with a non-monotonic transfer function
of reversed-wedge type that
have been investigated as an associative memory
\cite{Morita,Nishi,Ino} and a perceptron \cite{Boff,Monasson94}.
If a simple monotonic perceptron learns a rule from
examples presented by a non-monotonic perceptron,
the generalization error
remains non-vanishing even if an
infinite number of examples are presented by the teacher.
We study the limiting value and asymptotic behaviour
of the generalization error in such unlearnable cases.

This paper is composed of nine sections.
In the next section, the problem is formulated and the general
properties of generalization error are investigated.  Perceptron
and Hebbian learning algorithms in the on-line scheme are
investigated in section 3.
For each learning scheme, we calculate
the asymptotic behaviour of learning curve.
In section 4 we investigate
the effects of output noise on learning processes.
In section 5 we introduce the optimal learning rate
and calculate the
optimal generalization error.
The optimal learning rate obtained in section 5 contains an
unknown parameter for the student in
some contradiction to the idea of learning
because the learning process depends upon the unknown
teacher parameter.
Therefore, in section 6 we introduce
a learning rate independent of the
unknown parameter and optimize the rate to achieve a
faster convergence of generalization error.
In section 7, we allow the student to ask queries under
the Hebbian learning algorithm.  It is shown that learning
is accelerated considerably if the learning rate is
optimized.
In section 8 we optimize the learning dynamics
by a weight-decay term to avoid an
over-training problem in Hebbian learning
observed in section 3.
The last section contains summary and  discussions.
\section{Generic properties of generalization error}

Our problem is defined as follows.
The teacher signal is provided by a single-layer perceptron
with an $N$-dimensional weight vector
$\mbox{\boldmath $J^{0}$}$
and a non-monotonic (reversed-wedge) transfer function
\begin{equation}
T_{a}(v)={\rm sign}[v(a-v)(a+v)]
\label{Ta}
\end{equation}
where $v{\equiv}\sqrt{N}
(\mbox{\boldmath $J^{0}$}{\cdot}\mbox{\boldmath $x$})
/|\mbox{\boldmath $J^{0}$}|$, $\mbox{\boldmath $x$}$ is
the input vector normalized to unity,
$a$ is the width of the reversed wedge,
and ${\rm sign}$ denotes the sign function.
The student is a simple perceptron with the
weight vector $\mbox{\boldmath $J$}$ whose output is
\begin{equation}
S(u)={\rm sign}(u)
  \label{Su}
\end{equation}
where $u\,{\equiv}\,\sqrt{N}(\mbox{\boldmath $J$}
{\cdot}\mbox{\boldmath $x$})/|\mbox{\boldmath $J$}|$.
The components of
$\mbox{\boldmath $x$}$
are drawn independently from a uniform distribution
on the $N$-dimensional unit sphere.
The student can learn the rule of the teacher
perfectly if and only if $a=\infty$.

It is convenient to introduce the following two order parameters.
One is the overlap between $\mbox{\boldmath $J^{0}$}$
and $\mbox{\boldmath $J$}$
\begin{equation}
R=\frac{\mbox{\boldmath $J^{0}$}{\cdot}\mbox{\boldmath $J$}}
{|\mbox{\boldmath $J^{0}$}||\mbox{\boldmath $J$}|}
\label{q}
\end{equation}
and the other is the norm of the student weight vector
\begin{equation}
l=\frac{|\mbox{\boldmath $J$}|}{\sqrt{N}} .
  \label{l}
\end{equation}
In the limit $N\to\infty$ the random variables
$u$ and $v$ obey the normal distribution
\begin{equation}
P_{R}(u,v)=\frac{1}{2{\pi}\sqrt{1-R^{2}}}
\,{\exp}\left[-\frac{u^{2}+v^{2}-2Ruv}{2(1-R^{2})}\right] .
\label{Pq}
\end{equation}
The generalization error ${\epsilon}_{\rm g}$, or the student
probability of producing a wrong answer, can be obtained by
integrating the above distribution over the region satisfying
$T_{a}(v){\neq}S(u)$ in the two-dimensional $u$-$v$ space.
After simple calculations we find
\begin{equation}
{\epsilon}_{\rm g}\,{\equiv}\,E(R)=2\int_{a}^{\infty}
{\rm D}v\,H\left(\frac{-Rv}{\sqrt{1-R^{2}}}\right)
+2\int_{0}^{a}{\rm D}v\,H
  \left(\frac{Rv}{\sqrt{1-R^{2}}}\right)
\label{eg}
\end{equation}
where $H(x)=\int_{x}^{\infty}Dv$ and 
${\rm D}v\,{\equiv}\,\d v\,{\exp}(-v^{2}/2)/\sqrt{2\pi}$.

In figure 1 we plot $E(R)(={\epsilon}_{\rm g})$
for several values of the parameter $a$.
{}From this figure, we see that for $a=\infty$ (the learnable limit),
${\epsilon}_{\rm g}$ goes to zero when $R$ approaches $1$.
In contrast, for $a=0$, ${\epsilon}_{\rm g}$
goes to zero when $R$ reaches $-1$.
If $a$ is finite, the generalization error
shows highly non-trivial behaviour.
The critical value $R_{*}$ of the order parameter
is defined as the point where $E(R)$ is locally minimum.
Explicitly,
\begin{equation}
R_{*}=-\sqrt{\frac{2{\log}2-a^{2}}{2{\log}2}}
 \label{qs}
\end{equation}
which exists for $a\le a_{\rm c1}=\sqrt{2\log 2}=1.18$.
%
We plot in figure 2 the value of the global minimum 
of $E(R)$, the smallest possible
generalization error irrespective of learning algorithms.
In figure 3, we show the value of $R$ which gives the
global minimum.
We notice that
for $a<a_{\rm c2}\,{\equiv}\,0.80$,
$E_{\rm local}\equiv E(R=R_{*})$ is also
the global minimum, and
for $a>a_{\rm c2}$, the global minimum is $E(R=1)$.
Clearly the
optimal generalization error is obtained by training
the student weight vector $\mbox{\boldmath $J$}$
so that $R$ goes to 1 (or
$\mbox{\boldmath $J$}=\mbox{\boldmath $J^{0}$}$).
This critical value $a_{\rm c2}$ is given by the condition
$E(R=1)=E_{\rm local}$.

On the other hand, for $a<a_{\rm c2}$, the optimal
generalization cannot be achieved
even if the student succeeds in finding
$\mbox{\boldmath $J^{0}$}$ completely.
In this curious case, the optimal generalization
is obtained by training the student so that the student
finds his weight vector which satisfies
$R=R_{*}$ instead of $R=1$.
At $a=a_{\rm c2}$ the generalization error
has the maximum value as seen in figure 2.

\section{Dynamics of noiseless learning}
We now investigate the learning dynamics
with specific learning rules.
\subsection{Perceptron learning}

We first investigate the perceptron learning
\begin{equation}
\mbox{\boldmath $J$}^{m+1}=\mbox{\boldmath $J$}^{m}
-{\Theta}(-T_{a}(v)S(u))\,{\rm sign}(u)\,\mbox{\boldmath $x$}
  \label{Jm}
\end{equation}
where ${\Theta}$ is the step function and
$m$ stands for the discrete time step of
dynamics or the number of presented examples.
The standard procedure (see e.g. \cite{Kaba}) yields the rate of
changes of $l$ and $R$ in the limit $N\to\infty$ as
\begin{eqnarray}
\frac{\d l}{\d\alpha}=\frac{1}{l}\left[\frac{E(R)}{2}-F(R)l\right]
\label{dl} \\
\frac{\d R}{\d\alpha}=\frac{1}{l^2}
\left[-\frac{R}{2}E(R)+\left( F(R) R-G(R)\right) l\right]
\label{dq}
\end{eqnarray}
where $E(R)={\ll}1{\gg}_{R}$, $F(R)={\ll}u\,{\rm sign}(u){\gg}_{R}$
and $G(R)={\ll}v\,{\rm sign}(u){\gg}_{R}$.
The brackets ${\ll}{\cdots}{\gg}_{R}$
stand for the averaging with
respect to the distribution $P_R(u, v)$,
the integration being carried out over the region where
the student and the teacher give different outputs
$T_a(v)\ne S(u)$.
Hence the definition of $E(R)$ coincides with that of the
generalization error, $E(R)={\epsilon}_{\rm g}$,
as used in the previous section.
The other quantities $F(R)$ and $G(R)$ are evaluated in
a straightforward manner as
\begin{equation}
F(R)=-\frac{R}{\sqrt{2\pi}}(1-2{\Delta})+\frac{1}{\sqrt{2\pi}}
  \label{Fq}
\end{equation}
\begin{equation}
G(R)=-\frac{1}{\sqrt{2\pi}}(1-2{\Delta})+\frac{R}{\sqrt{2\pi}}
  \label{Gq}
\end{equation}
where ${\Delta}={\rm e}^{-a^{2}/2}$.
\subsubsection{Numerical analysis of differential equations}
We have numerically solved equations \eref{dl} and \eref{dq}.
The resulting flows of $R$ and $l$ are shown in figure 4
for $a=\infty$ under several initial conditions.
This figure indicates that $R$ reaches $1$
(perfect generalization state)
in the limit of ${\alpha}\,{\rightarrow}\,\infty$
and  $l\,{\rightarrow}\,\infty$ for any initial condition.
For finite ${\alpha}$, however,
behaviour of the flow strongly depends on the initial condition.
If we take a large $l$ as the initial value,  the
perfect generalization state $(R=1)$
is achieved after $l$ decreases at intermediate steps.
If we choose initial $R$ close to 1 and small $l$, the
perfect generalization is achieved after a decrease of $R$
is observed.
Similar phenomena have been reported
in the $K=2$ parity machine \cite{Kaba}.
Next we display the flows of $R$ and $l$
for unlearnable cases, for example,  $a=2.0$ 
in figure 5.  
There exists a stable and $a$-dependent fixed point $(R_{0},l_{0})$.
The generalization of the student halts at this fixed point
even if the flow of $R$ and $l$ starts from $R=1$ and large $l$.
\subsubsection{Asymptotic analysis of the learning curve}

When the rule is learnable ($a=\infty$), it is straightforward
to check the asymptotic behaviour 
${\epsilon}_{\rm g}=k\alpha^{-1/3}$, $k=\sqrt{2}(3\sqrt{2})^{-1/3}/\pi$,
from equations \eref{dl} and \eref{dq}.
When $a$ is finite, the fixed point value of $R$ is obtained from equations
\eref{dl} - \eref{Gq} as $R_{0}=1-2{\Delta}$.
Substituting this $R_{0}$ into $E(R)$,
we get the minimum value of the generalization
error $E_{0}={\epsilon}_{\rm min}(a)$ for perceptron learning.
In figures 2 and 3, we show $R_{0}$
and $E_{0}$ as functions of $a$.
Figure 2 indicates that the learning for
$a=a_{\rm c1}{\equiv}\sqrt{2{\log}2}$,
which is obtained from the condition $R_{0}=0$, is equivalent
to a random guess, ${\epsilon}_{\rm min}(a_{\rm c1})=0.5$.

Linearization of the right-hand side of equations \eref{dl} and \eref{dq}
around the fixed point yields the behaviour of the generalization
error near the fixed point.
Explicit expressions simplify when $a$
is large: it turns out that the
generalization error decays toward the minimum value
\begin{equation}
  E(R) \simeq 2H(a) \simeq 
    \frac{1}{\pi}{\Gamma}\left(\frac{1}{4}\right) {\Delta}^{3/4}
\end{equation}
exponentially as
$(\sqrt{2}/\pi) \exp ( -2 \Delta^{2/3}\alpha/\pi)$. 
\subsection{Hebbian learning}

In the Hebbian rule the dynamics of the student weight vector is
\begin{eqnarray}
\mbox{\boldmath $J$}^{m+1}=\mbox{\boldmath $J$}^{m}
+T_{a}(v)\,\mbox{\boldmath $x$} .
 \label{JmH}
\end{eqnarray}
This recursion relation of the $N$-dimensional vector
$\mbox{\boldmath $J$}$ is reduced to
the evolution equations of the order parameters as
\begin{eqnarray}
\frac{\d l}{\d\alpha}=\frac{1}{l}
  \left[\frac{1}{2}+\frac{2R}{\sqrt{2\pi}}
(1-2{\Delta})l\right]
  \label{dlH}\\
\frac{\d R}{\d\alpha}=\frac{1}{l^2}\left[
-\frac{R}{2}+\frac{2}{\sqrt{2\pi}}(1-2{\Delta})
(1-R^{2})l\right] .
 \label{dqH}
\end{eqnarray}
\subsubsection{Numerical analysis of differential equations}
In figure 6, we plot the flows in the $R$-$l$ plane
and the generalization error for $a=\infty$, $2.0$
and $a=0.5$.
We started the dynamics with the
initial condition $(R_{\rm init}, l_{\rm init})=(0.01,0.1)$.
This figure shows that $R$ reaches 1 for large $a$
and
$R$ approaches $-1$ for small $a$.
In order to find this bifurcation point near $R=0$,
we approximate equation \eref{dqH} around $R\,{\sim}\,0$ as
\begin{equation}
\frac{\d R}{\d\alpha}\,{\simeq}\,
  \frac{2}{\sqrt{2\pi}\,l}(1-2\Delta) .
 \label{dqdaH}
\end{equation}
If $a>a_{\rm c1}=\sqrt{2\log 2}=1.18$, the derivative
$\d R/\d \alpha$ is positive, and consequently
$R$ increases and eventually reaches $1$ in the limit
$\alpha{\rightarrow}\infty$.
If $a<a_{\rm c1}$, $R$ reaches $-1$
as $\alpha{\rightarrow}\infty$.
Figure 7 shows how the generalization error
behaves according to $a$.
For $a=0.5 (< a_{\rm c1})$,
${\epsilon}_{\rm g}$ has a minimum at some intermediate ${\alpha}$.
When the generalization error ${\epsilon}_{\rm g}$ passes
through this value, ${\epsilon}_{\rm g}$ begins to increase
toward the limiting value ${\epsilon}_{\rm min}(a)=1-2H(a)$.
Therefore, if the student learns excessively,
he cannot achieve the lowest generalization error
located at the global
minimum of $E(R)={\epsilon}_{\rm g}$ (over-training)
\cite{Biehl92,Opper}.

From figure 1 we see that $R$ must pass
through a local minimum of $E(R)$ at $R=R_{*}$ in order to
go to the state $R=-1$.
If the parameter $a$ satisfies $a<a_{\rm c2}=0.80$,
this local minimum is also the
global minimum. Therefore, if $a<a_{\rm c1}$,
although the generalization error decreases
until $R$ reaches $R_{*}$, it begins
to increase as soon as $R$ passes through the minimum
point $R=R_{*}$ and finally reaches a larger value at $R=-1$.

When the parameter $a$ lies in the range
$a_{\rm c2}<a<a_{\rm c1}$, the global minimum is located at $R=1$.
However, since $R$ goes to $-1$ for $ a<a_{\rm c1}$
(see equation \eref{dqdaH}), the generalization error increases
monotonically from $0.5$ (random guess) to $1-2H(a) (>0.5)$
for the parameter range $a_{\rm c2}< a<a_{\rm c1}$.
We can regard this as a special case of
over-training.  We conclude that
over-training appears for all $a<a_{\rm c1}$.
\subsubsection{Asymptotic analysis of the learning curve}
With the same technique as in the previous section,
we obtain the asymptotic form of the generalization error
when $a=\infty$
in the limit $\alpha{\rightarrow}\infty$ as
\begin{equation}
{\epsilon}_{\rm g}=\frac{1}{\sqrt{2\pi}}\,\frac{1}{\sqrt{\alpha}}
  \label{egb}
\end{equation}
which is a well-known result \cite{Vallet}.
For finite $a$ satisfying $a>a_{\rm c1}$,
simple manipulations as before show that 
the stable fixed point is at $R=1$ and
the differential equations \eref{dlH} and \eref{dqH}
yield the asymptotic form of the generalization error as
\begin{equation}
{\epsilon}_{\rm g}=\frac{1}{\sqrt{2\pi}(1-2{\Delta})}
\frac{1}{\sqrt{\alpha}}+2H(a) .
 \label{egc}
\end{equation}
The limiting value $2H(a)$ is the best possible value obtained
in section 2.
On the other hand, for $a<a_{\rm c1}$,
\begin{eqnarray}
{\epsilon}_{\rm g}=\frac{1}{\sqrt{6\pi}(1-2{\Delta})}
\frac{1}{\sqrt{\alpha}}
 +1-2H(a).
 \label{egd}
\end{eqnarray}
%
The rate of approach to the asymptotic value, $1/\sqrt{\alpha}$,
in equations \eref{egc} and \eref{egd} agrees with the
corresponding behaviour in the Gibbs learning of unlearnable
rules \cite{Kim}.
\section{Learning under output noise in the teacher signal}
We now consider the situation where
the output of the teacher
is inverted randomly with a rate $\lambda\,({\leq}1/2)$
for each example.
We show that the parameter $a$ plays
essentially the same role as output noise in the teacher signal.
\subsection{Perceptron learning}

According to references \cite{Kaba,Bar,Biehl95},
the effect of output noise is taken into account
in the differential equations (\ref{dl}) and (\ref{dq}) by
replacing $E(R)$, $F(R)$ and $G(R)$
with $\tilde{E}_{\lambda}(R)$, $\tilde{F}_{\lambda}(R)$
and $\tilde{G}_{\lambda}(R)$ as follows
\begin{equation}
 \eqalign{
\tilde{E}_{\lambda}(R)  =  (1-\lambda)E(R)+{\lambda}E^{c}(R) \\
\tilde{F}_{\lambda}(R)  =  (1-\lambda)F(R)+{\lambda}F^{c}(R) \\
\tilde{G}_{\lambda}(R)  =  (1-\lambda)G(R)+{\lambda}G^{c}(R)
\label{Flq}}
\end{equation}
where $E^{c}$, $F^{c}$ and $G^{c}$ 
correspond to $E$, $F$ and $G$, the only difference being
that the integration is over the region
satisfying $T_{a}(v)=S(u)$.
%

We study  the
asymptotic behaviour of the learning
curve in the limit of small noise level ${\lambda}{\ll}1$.
For the learnable case $a=\infty$, equations \eref{dl} and \eref{dq}
with \eref{Flq} taken into account have the fixed point at
$R=R_0\equiv 1-2\lambda$, $l=l_0\equiv (2\sqrt{2\pi\lambda})^{-1}$ for
$\lambda \ll 1$.
Linearization around this fixed point leads to the asymptotic behaviour
\begin{equation}
\eqalign{
l  {\sim} l_{0}\left
[1+{\cal O}({\rm e}^{-8{\lambda}^{3/2}{\alpha}})\right] \\
1-R  {\sim}  (1-{R}_{0})\left[
1+{\cal O}({\rm e}^{-8{\lambda}^{3/2}{\alpha}})\right] .
\label{Le0}
}
\end{equation}
Therefore, the generalization error ${\epsilon}_{\rm g}$
converges to a finite value $E(R=1-2\lambda )=2{\lambda}^{1/2}/\pi$
exponentially, ${\exp}(-8{\lambda}^{3/2}{\alpha})$.

According to Biehl {\etal}\cite{Biehl95},
it is useful to distinguish two
performance measures of on-line learning, the
generalization error ${\epsilon}_{\rm g}$ and the
prediction error ${\epsilon}_{\rm p}$.
The generalization error ${\epsilon}_{\rm g}$ is the probability
of disagreement between the student and the genuine
rule of the teacher as we have discussed.
On the other hand,
the prediction error ${\epsilon}_{\rm p}$
is the probability for disagreement between
the student and the noisy teacher output for an arbitrary input.
In the present case, the prediction error
${\epsilon}_{\rm p}$ and generalization error
${\epsilon}_{\rm g}$ satisfy the relation
\begin{equation}
{\epsilon}_{\rm p}={\lambda}+(1-2{\lambda}){\epsilon}_{\rm g} .
 \label{ep}
\end{equation}

For the unlearnable case of large but finite $a$ under
small noise level, the fixed point value of $R$ is found to
be $R_0(\lambda)=(1-2\Delta)(1-2\lambda)$.
The expression of the fixed point $l_0(\lambda)$ is too complicated
 and is omitted here.
Linearization near this fixed point shows that 
the generalization error converges to
$(2/\pi){\lambda}^{1/2}+2H(a)$ exponentially as
${\exp}(-t_{-}{\alpha})$ for
large $a$ and small ${\lambda}$,
where
\begin{equation}
t_{-}=\frac{(-8{\lambda}^{3/2}-2{\lambda}^{1/2})
-\,\sqrt{(-8{\lambda}^{3/2}+2{\lambda}^{1/2})^{2}-(8{\Delta}
+4{\lambda}^{-1}{\Delta}^{2})}}{2} .
\label{tpm}
\end{equation}
The prediction error is given by
${\epsilon}_{\rm p}=\lambda+(1-2\lambda){\epsilon}_{\rm g}$.
\subsection{Hebbian learning}

The differential equations of the order parameters
for noisy Hebbian learning are
\begin{eqnarray}
\frac{\d l}{\d\alpha} = \frac{1}{l}\left[\frac{1}{2}
+\frac{2R}{\sqrt{2\pi}}(1-2\Delta)(1-2\lambda)l\right]
  \label{dlda32}\\
\frac{\d R}{\d\alpha} = \frac{1}{l^2}\left[
-\frac{R}{2}+\frac{2}{\sqrt{2\pi}}
(1-2\Delta)(1-2\lambda)(1-R^{2})l\right] .
  \label{dqda32}
\end{eqnarray}
We plot the generalization error for $a=0.5$ 
in figure 8
by solving these differential equations numerically.
We saw in the previous section that
the over-training appears in the absence of noise
if $a<a_{\rm c1}=\sqrt{2{\log}2}$, which is also the case
when there is small noise (e.g.
$\lambda =0.01$). 
For larger $\lambda$ (e.g. $\lambda =0.20$),
however, there appears no minimum in ${\epsilon}_{\rm g}$ as
$\alpha$ increases.
This implies in terms of figure 1 that $R$ becomes
stuck at an intermediate $R$ before it reaches $R_{*}$.

The asymptotic form for the noisy case can be derived simply by
replacing $(1-2\Delta)$ in the asymptotic form of the
noiseless case with $(1-2\Delta)(1-2\lambda)$.
Thus
$\Delta={\rm e}^{-a^{2}/2}$ and $\lambda$
have the same effect on the asymptotic
generalization ability.
A similar effect is reported for the non-monotonic Hopfield
model \cite{Nishi,Ino} which works as an associative memory.
If we embed patterns by the Hebb rule in the network,
the capacity of the network drastically deteriorates for small $a$.
\section{Optimization of learning rate}

We have so far investigated the learning processes
with a fixed learning rate.
In this section we consider optimization of the learning rate
to improve the learning performance.
It turns out that the perceptron learning with optimized
learning rate achieves the best possible generalization error
in the range $a\ge a_{\rm c1}$.

We first introduce the learning rate
$g(\alpha)$ in our dynamics.
As an example, the learning dynamics for the perceptron algorithm
is written as
\begin{equation}
\mbox{\boldmath $J$}^{m+1} =
\mbox{\boldmath $J$}^{m} -
g(\alpha)\,{\Theta}(-T_{a}(v)S(u))\,{\rm sign}(u)\,
\mbox{\boldmath $x$} .
  \label{Jm4}
\end{equation}
This optimization procedure is different
from the technique of Kinouchi and Caticha \cite{Kino}.
They investigated the on-line dynamics with a
general weight function $f(T_{a}(v),u)$ as
\begin{equation}
\mbox{\boldmath $J$}^{m+1}=\mbox{\boldmath $J$}^{m}
+f(T_{a}(v),u)\,T_{a}(v)\,\mbox{\boldmath $x$}
  \label{Jm41}
\end{equation}
and chose $f(T_{a},u)$ so that it maximizes the
increase of $R$ per learning step.
In contrast, our optimization procedure adjusts the parameter
$g(\alpha)$ keeping the learning algorithm unchanged.

\subsection{Perceptron learning}
\subsubsection{Trajectory in the $R$-$l$ plane}

The trajectories
in the $R$-$l$ plane can be derived explicitly
for the optimal learning rate $g_{\rm opt}(\alpha)$.
The differential
equations with the learning rate $g(\alpha)$ are
\begin{eqnarray}
\frac{\d l}{\d\alpha}=\frac{g(\alpha)^{2}
E(R)/2-g(\alpha)F(R)l}{l}
  \label{dlda41}\\
\frac{\d R}{\d\alpha}=\frac{-R
E(R)g(\alpha)^{2}/2+g(\alpha)\left[F(R)R-G(R)\right]l}{l^{2}}
{\equiv}L(g(\alpha)) .
  \label{dqda41}
\end{eqnarray}
Now we choose the parameter $g$ to
maximize $L(g(\alpha))$ with the aim
to accelerate the increase of $R$
\begin{equation}
g_{\rm opt}(\alpha)=\frac{\left[F(R)\,R-G(R)\right] l}{RE(R)} .
  \label{ga}
\end{equation}
Substituting this $g$ into
equations (\ref{dlda41}) and (\ref{dqda41})
and taking their ratio, we find
\begin{equation}
\frac{\d R}{\d l}=
-\frac{\left[F(R)\,R-G(R)\right]\,R}{\left[F(R)\,R+G(R)\right]l} .
  \label{dqda43}
\end{equation}
Using equations (\ref{Fq}) and (\ref{Gq}) we obtain the trajectory
in the $R$-$l$ plane as
\begin{equation}
(1+R)^{-(1+A)/A}(1-R)^{(1-A)/A}R=c\,l
  \label{ql1}
\end{equation}
where $A=1-2\Delta$ and $c$ is a constant.

In figures 9 and 10, we plot the above
trajectory for $a=2.0$ and $0.5$, respectively,
by adjusting $c$ to reproduce the initial conditions
$(R_{\rm init},l_{\rm init})=(0.01, 0.10),
(0.01, 1.00)$ and $(0.01, 2.00)$.
These figures indicate that the student
goes to the state of $R=1$ after
infinite learning steps $(\alpha{\rightarrow}\infty)$ for any
initial condition.
The final value of $l$ depends on $a$.
If $a$ is small (e.g., 0.5), $l$ increases indefinitely as
$\alpha\to\infty$.
On the other hand, for larger $a$,
$l$ is seen to decrease as $\alpha$
goes to $\infty$.
We investigate this $a$-dependence of $l$ in more detail in
the next subsection.

We plot the corresponding generalization error in
figures 11 and 12.
We see that for $a=2.0$,
the generalization ability is improved significantly.
However, for $a=0.5$, the generalization ability becomes
worse than that for $g=1$ (the unoptimized case).

We note that the above optimal learning rate $g_{\rm opt}(\alpha)$
contains the parameter $a$ unknown to the student.
Thus this choice of $g(\alpha)$ is not perfectly consistent with
the principles of supervised learning.
We will propose an improvement
on this point in section 6 using a parameter-free learning rate.
For the moment, we may take the result of the present section
as a theoretical estimate of the best possible optimization result.

\subsubsection{Asymptotic analysis of the learning curve}

Let us first investigate the learnable case.
The asymptotic forms of $R$,
$l$, ${\epsilon}_{\rm g}$ and $g$ as $R\to 1$
are obtained from the same analysis
as in the previous section as $R=1-8/\alpha^2$,
$l  =  c\,{\rm e}^{-16/\alpha^{2}}$
and
\begin{eqnarray}
{\epsilon}_{\rm g} = \frac{4}{\pi\alpha}
 \label{eg41}\\
g(\alpha)= 2\sqrt{2\pi}\,\frac{l}{\alpha}
 =2c\sqrt{2\pi}\frac{{\rm e}^{-16/\alpha^{2}}}{\alpha}
\label{ga41}
\end{eqnarray}
where $c$ is a constant depending on the initial condition.
The decay rate to vanishing generalization error
is improved from ${\alpha}^{-1/3}$
for the unoptimized case \cite{Bar}
to ${\alpha}^{-1}$.
This ${\alpha}^{-1}$-law is the same as in
the off-line (or batch) learning \cite{Opper90}.
We also  see that $l$ approaches $c$ as $R$ reaches $1$.

We next investigate the unlearnable case ${\Delta}{\neq}0$.
The asymptotic forms are
\begin{equation}
\eqalign{
R =  1-\frac{2{\pi}H(a)}{
(1-2{\Delta})^{2}}\frac{1}{\alpha} \\
l =  c\,{\alpha}^{-2{\Delta}/(1-2{\Delta})}
\label{el42}}
\end{equation}
\begin{equation}
{\epsilon}_{\rm g}= \frac{\sqrt{2}}{\pi}
\frac{\sqrt{2{\pi}H(a)}}{1-2{\Delta}}
\frac{1}{\sqrt{\alpha}}+2H(a)
  \label{eg42}
\end{equation}
and the optimal learning rate $g_{\rm opt}$ is
\begin{equation}
g_{\rm opt}(\alpha)\,{\simeq}\,
c\frac{\sqrt{2\pi}}{1-2\Delta}
\frac{{\alpha}^{-2{\Delta}/(1-2{\Delta})}}
{\alpha} .
  \label{ga42}
\end{equation}
{}From the asymptotic form of $l$, we find that $l$ diverges with
$\alpha$ for $a<a_{\rm c1}=\sqrt{2{\log}2}$
and goes to zero
for $a>a_{\rm c1}$ as observed in the previous subsection.
It is interesting that, for $a$ exactly equal to $a_{\rm c1}$,
$g_{\rm opt}$ vanishes and
the present type of optimization does
not make sense.

For $a>a_{\rm c2}=0.80$, the generalization error
converges to the optimal value $2H(a)$ as
${\alpha}^{-1/2}$.
This is the same exponent as that of the
Hebbian learning as we saw in the previous section.
For $a<a_{\rm c2}$, in order to get the optimal
overlap $R=R_{*}$,
we must stop the on-line dynamics
before the system reaches the state $R=-1$.
Accordingly, the method discussed in this
section is not useful for the purpose of improvement
of generalization ability for $a<a_{\rm c2}$.

\subsection{Hebbian learning}

The Hebbian learning with learning rate $g(\alpha )$ is
\begin{eqnarray}
\mbox{\boldmath $J$}^{m+1}=\mbox{\boldmath $J$}^{m}
+g(\alpha ) T_{a}(v)\,\mbox{\boldmath $x$} .
 \label{JmHg}
\end{eqnarray}
%
Using the same technique as in the previous subsection, we
find the optimal learning rate for the
Hebbian learning $g_{\rm opt}^{\rm H}(\alpha)$ as
\begin{equation}
g_{\rm opt}^{\rm H}(\alpha)=\sqrt{\frac{2}{\pi}}
\frac{(1-2{\Delta})(1-R^2)l}{R}.
 \label{go1}
\end{equation}
The $R$-$l$ trajectory is
\begin{equation}
\frac{R}{(1-R^{2})}=c\,l
  \label{ql2}
\end{equation}
where $c$ is a constant determined by the initial condition.
It is very interesting that this trajectory is independent of $a$.

The asymptotic forms of various quantities for
$a>a_{\rm c1}$ of the Hebbian learning are
\begin{equation}
\eqalign{
R =  1-\frac{\pi}{4(1-2{\Delta})^{2}}
\,\frac{1}{\alpha} \\
l  =  c\,{\alpha}
\label{el43}}
\end{equation}
and
\begin{eqnarray}
{\epsilon}_{\rm g} = \frac{1}{\sqrt{2\pi}(1-2{\Delta})}
\frac{1}{\sqrt{\alpha}} + 2H(a)
  \label{eg43}\\
g(\alpha) = c .   \label{ga43}
\end{eqnarray}
Accordingly, for $a>a_{\rm c1}$, the asymptotic form of
the generalization error is the same as for $g=1$.
However, in the parameter region
$a<a_{\rm c1}$, the generalization ability
deteriorates by introducing the
optimal learning rate if we select
an initial condition satisfying $R>0$.
To see this, we note that $\d R/\d {\alpha}$
is approximated around $R=0$ as 
$\d R/\d {\alpha}\,{\simeq}\,2(1-2{\Delta})^{2}/{\pi}R$ 
with using $g_{\rm opt}^{\rm H}$.
Therefore if
we start the learning dynamics from $R>0$, the
overlap $R$ goes to $1$ and the generalization error
approaches $2H(a)$ which is not acceptable at all because
it exceeds 0.5.
On the other hand, for $a<a_{\rm c1}$ and $R_{\rm init}<0$, the generalization
error approaches $1-2H(a)$ (less than 0.5 but not optimal) as
\begin{equation}
{\epsilon}_{\rm g}=\frac{1}{\sqrt{2\pi}(1-2{\Delta})}
\frac{1}{\sqrt{\alpha}} + 1-2H(a) .
  \label{eg44}
\end{equation}
Thus an over-training appears.
We must notice that the
prefactor of the generalization error changes from $1/\sqrt{6\pi}$
in equation \eref{egd}
to $1/\sqrt{2\pi}$ in equation \eref{eg44} by
introducing the optimal learning rate.
Therefore the optimization by using the learning
rate $g(\alpha)$ is not very useful
for the Hebbian learning.
\section{Optimal learning without unknown parameters}

As we mentioned in section 5, the generalization error
obtained there is the theoretical (not practical)
lower bound because the
optimal learning rate $g_{\rm opt}$
contains a parameter $a$ unknown
to the student.
In this section we propose a method to avoid this difficulty
for the perceptron learning algorithm.

For the learnable case we choose the learning rate $g$ as
\begin{equation}
g=\frac{k}{\alpha}\,l
  \label{go61}
\end{equation}
which is nothing but the asymptotic form (\ref{ga41}) of the
previous optimized learning rate.
Substituting this into equation \eref{dqda41} with \eref{ga},
we find $R=1-8/\alpha^2$ when $R$ is close to unity and
correspondingly
\begin{equation}
{\epsilon}_{\rm g}=\frac{4}{\pi\alpha}
  \label{eg60}
\end{equation}
which agrees with the result of Barkai {\etal}\cite{Bar}.

For the unlearnable case,
we  assume $g(\alpha)=kl/\alpha$ as before and 
find the general solution for $R=1-\varepsilon$ as
\begin{equation}
{\varepsilon}=\frac{k^{2}H(a)}{bk-1}
\frac{1}{\alpha}
+A\left(
\frac{k}{\alpha}
\right)^{bk}
\label{solep1}
\end{equation}
where $b\,{\equiv}\,\sqrt{2/\pi}(1-2{\Delta})$.
The first term dominates asymptotically if $bk>1$.
In this case, we have
\begin{equation}
{\epsilon}_{\rm g}=2H(a)+
\sqrt{\frac{2k^2 H(a)}{bk-1}} \frac{1}{\pi\sqrt{\alpha}} .
\label{kka}
\end{equation}
The second term on the right-hand side is minimized by
choosing
\begin{equation}
k=\frac{\sqrt{2\pi}}{1-2{\Delta}} 
  \label{k61}
\end{equation}
which satisfies $bk>1$ as required.  Equation \eref{kka} makes
sense for $\Delta >2\sqrt{\log 2}$ if $k$ is chosen as above.

When $bk<1$, the asymptotic form of the generalization error is
\begin{equation}
{\epsilon}_{\rm g}=2H(a)+\frac{\sqrt{2A}}{\pi}
\left(
\frac{\sqrt{2\pi}}{\alpha}
\right)^{bk/2} .
\label{eps12}
\end{equation}
This formula is valid for $b>0$ or $a<a_{\rm c1}$.  
Similar crossover between two types of asymptotic forms
was reported in the problem of one-dimensional
decision boundary \cite{Kaba95}.
\section{Hebbian learning with queries}
We have so far assumed that the student is 
trained using examples drawn from a uniform distribution
on the $N$-dimensional sphere $S^{N}$. 
It is known for the learnable case \cite{Kinzel} that
selecting training examples out of a limited set
sometimes improves the performance of learning.
We therefore investigate in the present section how the method
of Kinzel and Ruj\'{a}n \cite{Kinzel} works for an unlearnable
rule.
\subsection{Learning with queries under fixed learning rate}
The learning dynamics we choose here is nothing but the Hebbian algorithm
\eref{JmH}.
In section 3, the student was trained by inputs 
$\mbox{\boldmath $x$}$ uniform on $S^{N}$.
In the present section we follow reference \cite{Kinzel} and 
use selected inputs which lie on the  borderline, 
$\mbox{\boldmath $J$}{\cdot}
\mbox{\boldmath $x$}=0$ or $u=0$, at every dynamical step.
The idea behind this choice is that the student is not
confident for inputs just on the decision boundary and thus
teacher signals for such examples should be more
useful than generic inputs.

We use the following  conditional distribution,
instead of $P_{R}(u,v)$ in equation \eref{Pq}, 
in order to get the differential equations
\begin{equation}
P_{R}(v|u=0)=\sqrt{2\pi}{\delta}(u)P_{R}(u,v) .
\label{Hq2}
\end{equation}
Using this distribution, 
we obtain the next differential equations
\begin{equation}
\frac{\d l^{2}}{\d\alpha}=1
\label{Hq3}
\end{equation}
\begin{equation}
\frac{\d R}{\d \alpha}=\frac{1}{l}\left[
\sqrt{\frac{2}{\pi}}\sqrt{1-R^{2}}
\left\{
1-2\, {\exp}\left(-\frac{a^{2}}{2(1-R^{2})}\right)
\right\}
-\frac{R}{2l}
\right] .
\label{Hq4}
\end{equation}
In figure 13, we plotted the generalization error 
for $a=1.0$ 
by numerical integration of the above differential equations. 
We see that  the 
generalization ability of student is improved 
and the problem of over-training is avoided.

In order to investigate the asymptotic form of the
generalization error, 
we solve the differential equations in the limit of 
${\alpha}{\rightarrow}\infty$. 
Equation \eref{Hq3} can be solved easily as 
$l=\sqrt{\alpha}$.
For the learnable case $a{\rightarrow}\infty$,
using $R=1-{\varepsilon}$ and  
${\varepsilon}{\rightarrow}0$, we obtain 
${\varepsilon}={\pi}/(16\alpha )$ and  
the generalization error as
\begin{equation}
{\epsilon}_{\rm g}=\frac{1}{2\sqrt{2\pi}}
\frac{1}{\sqrt{\alpha}} .
\label{Hq7}
\end{equation}
The numerical prefactor has been reduced by
a half from equation \eref{egb}. 

For finite $a$, equation \eref{Hq4} has fixed points at
$R_0=\pm 1$ and
\begin{equation}
R_{1}^{(\pm )}={\pm}\sqrt{\frac{2{\log}2-a^{2}}{2{\log}2}} .
\label{Hq9}
\end{equation}
The latter fixed point exists only for $a<a_{\rm c1}=\sqrt{2\log 2}$.
Thus, if $a>a_{\rm c1}$, $|R|$ eventually approaches 1, and the
exponential term in equation \eref{Hq4} can be neglected.
This implies that the asymptotic analysis for the learnable
case applies without modification.
The resulting asymptotic form of the generalization error is
\begin{equation}
{\epsilon}_{\rm g}=\frac{1}{2\sqrt{2\pi}}\frac{1}{\sqrt{\alpha}}
+2H(a) .
\label{Hq8}
\end{equation}

If  $a<a_{\rm c1}$, the system is attracted to the fixed
point $R_1^{(-)}$ according to the expansion of the right-hand
side of equation \eref{Hq4} around $R=0$,
\begin{equation}
\frac{\d R}{\d \alpha}\,{\simeq}\,
\frac{1}{l}\sqrt{\frac{2}{\pi}}(1-2{\Delta}) 
\label{Hq10}
\end{equation}
which is negative if $a<a_{\rm c1}$.
It is remarkable that $R_1^{(-)}$ coincides with $R_{*}$
which gives the global minimum of $E(R)$ for
$a<a_{\rm c2}=0.80$.
Therefore, for $a<a_{\rm c2}$,
the present Hebbian learning with queries achieves
the best possible generalization error.
In the range $a_{\rm c2}<a<a_{\rm c1}$, $R=R_1^{(-)}=R_{*}$
is not the global minimum of $E(R)$ but is only a local minimum.
However, as seen in figure 13, over-training has disappeared
in this region by introducing queries.

The asymptotic behaviour for $a<a_{\rm c1}$ is found to be
\begin{eqnarray}
{\epsilon}_{\rm g}=
{\epsilon}_{\rm opt}
-\frac{16{\log}2\sqrt{2{\log}2-a^{2}}}
{a^{2}}\left[
1-Q(2,\frac{1}{2}{\log}2)\right]
\nonumber\\
\times{\exp}\left[
-\frac{8{\log}2}{\sqrt{\pi}a}
\sqrt{2{\log}2-a^{2}}
\sqrt{\alpha}
\right]
\label{Hq13}
\end{eqnarray}
where $Q(x,y)$ is the incomplete gamma function and
the asymptotic value 
${\epsilon}_{\rm opt}=E(R_{*})$ is optimal for $a < a_{\rm c1}$.

\subsection{Optimized Hebbian learning with queries}
We next introduce the parameter $g$ into the
Hebbian learning with queries and optimize $g$ so that 
$R$ goes to $1$ as quickly as possible. 
As discussed in section 5, this strategy 
works only for $a>a_{\rm c2}$ since $R=1$ is not the optimal
value if $a<a_{\rm c2}$.
Using the same technique as section 5, we find the optimal 
learning rate as 
\begin{equation}
g_{\rm opt}=\frac{l}{R}\sqrt{\frac{2}{\pi}}
\sqrt{1-R^{2}}\left\{
1-2\,{\exp}\left(-\frac{a^{2}}{1-R^{2}}\right)
\right\} .
\label{Hq17}
\end{equation}
For the learnable case, the solution for $R$ is 
\begin{equation}
R=\sqrt{1-c\,{\exp}(-\frac{2\alpha}{\pi})} 
\label{Hq19}
\end{equation}
where $c$ is a constant.
The generalization error decays to 
zero as 
\begin{equation}
{\epsilon}_{\rm g}=\frac{\sqrt{c}}{\pi}
{\exp}(-\frac{\alpha}{\pi}) 
\label{Hq20}
\end{equation}
where $c$ is determined by the initial condition. 
This exponential decrease for the learnable case is
in agreement with reference \cite{Kino} where
the optimization of the type of equation \eref{Jm41}
was used together with queries.
The asymptotic forms of the order parameter $l$ and 
optimal learning rate $g_{\rm opt}$ are
\begin{equation}
l=c'\sqrt{1-c\,{\exp}\,(-\frac{2\alpha}{\pi})}
\label{Hq21}
\end{equation}
\begin{equation}
g_{\rm opt}(\alpha)=
c'\sqrt{\frac{2c}{\pi}}\,{\exp}\,(-\frac{\alpha}{\pi})
\label{Hq22}
\end{equation}
where $c'$ is determined by the initial condition. 

Next we investigate the case of finite $a$. 
Using the same asymptotic analysis as in the learnable case, 
we obtain the asymptotic form of generalization error 
${\epsilon}_{\rm g}$ as 
\begin{equation}
{\epsilon}_{\rm g}=2H(a)+\frac{\sqrt{c}}{\pi}\,{\exp}\,
(-\frac{\alpha}{\pi}) .
\label{Hq23}
\end{equation}
The limiting value $2H(a)$ is the theoretical lower bound 
for $a>a_{\rm c2}=0.80$.
We therefore have found a method of optimization 
to achieve the best possible
generalization error with a very fast, exponential, asymptotic
approach for $a>a_{\rm c2}$. 
The present method of optimization
does not work appropriately for $a<a_{\rm c2}$
because $R=1$, to which the present method is designed to force
the system, is not the best value of $R$ in this range of $a$.

It is worth investigating whether 
the exponent of decay changes or not by using a parameter-free
optimal learning rate as in 
section 7. 
If $a>a_{\rm c1}$, there exists only one 
fixed point $R=1$.
Therefore, the $a$-dependent term ${\exp}(-a^{2}/(1-R^{2}))$ 
in equation \eref{Hq17}
does not affect the asymptotic analysis. 
We may therefore conclude that the asymptotic 
form of generalization error does not change 
by optimal learning rate without the unknown parameter $a$.
\section{Avoiding over-training by a weight-decay term}

We showed in section 3 that the over-training
appears for the unlearnable case $a<a_{\rm c1}$
by the Hebbian learning.
If $a<a_{\rm c1}$, the flow of $R$ goes to
$-1$ for any initial condition passing through
the local minimum of $E(R)$ at $R=R_{*}$.
Consequently, the generalization ability of the student
decreases as he learns excessively.
In order to avoid this difficulty,
we must stop the dynamics on the way to the state $R=-1$.
For this purpose, we may use
the on-line dynamics with a
weight-decay term or a forgetting term \cite{Biehl92}.

 The on-line dynamics by the Hebbian rule is modified with the
weight-decay term as
\begin{equation}
\mbox{\boldmath $J$}^{m+1}=(1-\frac{\Lambda}{N})
\mbox{\boldmath $J$}^{m}+T_{a}(v)\,\mbox{\boldmath $x$} .
  \label{Jm71}
\end{equation}
The fixed point of the above dynamics is
\begin{equation}
R_{0}=\frac{2(1-2\Delta)}
{\sqrt{\pi{\Lambda}+4(1-2{\Delta})^{2}}}.
 \label{q071}
\end{equation}
In order to get the optimal value, we choose $R_{0}$ so that
it agrees with $R_{*}$ which gives the global
minimum of $E(R)$ for $a<a_{\rm c1}$.
{}From this condition,
 we obtain the optimal ${\Lambda}_{\rm opt}$ as
\begin{equation}
{\Lambda}_{\rm opt}=\frac{4a^{2}(1-2\Delta)^{2}}
{\pi(2{\log}2-a^{2})}.
  \label{L071}
\end{equation}
Using this $\Lambda_{\rm opt}$, we solve the
differential equations numerically and
plot the result in figure 14 for $a=0.5 (< a_{\rm c1})$.
We see that the over-training disappears
and the generalization error converges to the optimal value.

We next investigate how fast this convergence is achieved.
For this purpose, we linearize the differential equations around
the fixed point
to obtain
\begin{equation}
1-R{\sim}(1-R_{0})
\left\{ 1+{\cal O}\left[ {\exp}
(-2a^{2}(1-2\Delta)^{2}\left(
\frac{\pi(2{\log}2-a^{2})+4}
{\pi(2{\log}2-a^{2})}\right) {\alpha})\right]\right\}.
\label{e7}
\end{equation}
We warn here that $\Lambda_{\rm opt}$ in equation
\eref{L071} depends on
$a$ which is unknown to the student.
Therefore, the result obtained in this section
gives the theoretical upper bound of the generalization ability.

\section{Summary and Discussions}

We have analyzed the problem of on-line learning by the perceptron
and Hebbian algorithms.
For the unlearnable case, the
generalization error decays exponentially to a finite value
$E(R_{0})$ with $R_{0}=1-2{\Delta}$
in the case of the perceptron learning.
For the Hebbian learning, the generalization error
decays to $2H(a)$, the best possible value,
for $a>a_{\rm c1}$ and to $1-2H(a)$
for $a<a_{\rm c1}$, both proportionally to $\alpha^{-1/2}$.
In this latter parameter region $a<a_{\rm c1}$,
we observed the phenomenon of over-training.

We also investigated the learning under output noise.
For the learnable case of the perceptron algorithm, the
order parameters $R$ and $l$ are attracted toward a fixed point
$(R_{0},l_{0})$ asymptotically with an exponential law.
As a result, the generalization error decays to
a finite value exponentially.
On the other hand,
for the unlearnable case of the perceptron learning,
the generalization error decays exponentially
to a finite value $E((1-2{\Delta})(1-2{\lambda}))$.
For the Hebbian learning, the generalization error
decays to $2H(a)$ in proportion to $1/\sqrt{\alpha}$
for $a>a_{\rm c1}$ and to $1-2H(a)$ with
also proportionally to $1/\sqrt{\alpha}$
for $a<a_{\rm c1}$.

We introduced the learning rate $g(\alpha)$ in on-line dynamics
and optimized it to maximize $\d R/\d\alpha$.
By this treatment we obtained a closed form trajectory of
$R$ and $l$. The generalization ability of
the student has been shown to increase for
$a>a_{\rm c2}=0.80$ in the case of the perceptron learning
algorithm.
For the unlearnable case, the
generalization error decays to the best possible value
$2H(a)$ in proportion to
$1/\sqrt{\alpha}$.
For the Hebbian learning, 
the asymptotic generalization
ability did not change by this optimization procedure.

Unfortunately, in the parameter range $a<a_{\rm c2}$,
we found it impossible to
obtain an optimal performance
for the perceptron learning within our
procedure of optimization.
To overcome this difficulty,
we investigated the on-line dynamics
with a weight-decay term for the Hebbian learning.
Using this method, we could eliminate
the over-training, and the generalization error
converges to the optimal value exponentially.

We also introduced a new learning rate independent of the
unknown parameter $a$.
We assumed $g(\alpha)=kl/\alpha$ and optimized
$k$ so that the generalization error decays to
the minimum value as quickly as possible.
As a result, for the unlearnable case of
$a>a_{\rm c1}$ the prefactor was somewhat improved
although the exponent of decay did not change.

The Hebbian learning with queries was also investigated. 
If the student is trained by the Hebbian algorithm 
using inputs on the decision boundary, 
his generalization ability is improved 
except in the range $a_{\rm c2} < a < a_{\rm c1}$. 
This is a highly non-trivial result because
this choice of query works well 
for the unlearnable case where student does not know the
structure of the teacher. 
We next introduced the optimal learning 
rate in the on-line Hebbian learning with queries 
and obtained  very fast convergence of generalization error. 
For $a >a_{\rm c1}$, the generalization error 
converges to its optimal value exponentially. 

We have observed exponential decays to limiting values in
various situations of unlearnable rules.  This fast convergence
may originate in the large size of the asymptotic space;
if the liming value of $R$ is unity, only a single point in
the $\mbox{\boldmath $J$}$-space, 
$\mbox{\boldmath $J$}=\mbox{\boldmath $J^{0}$}$,
is the correct destination of learning dynamics, a
very difficult task. 
If, on the other hand, $R$ approaches $R_0 (<1)$, there
are a continuous number of allowed student vectors,
and to find one of these should be a relatively easy
process, leading to exponential convergence.

\ack
The authors gratefully acknowledge useful discussions with 
Professor Shun-ichi Amari.

\Figures
\Figure{
Generalization error as a function of the overlap $R$ for
$a=\infty$, $2.0$, $1.0$, $0.5$ and
$0$.
For $a=\infty$, the generalization
error decreases to zero as $R$ goes to $1$.
For $a=0$, the
generalization error decays to zero as
$R$ goes to $-1$ instead of $1$.
}
\Figure{
The global minimum value
of $E(R)$ which corresponds to the optimal
value of the generalization error ${\epsilon}_{\rm opt}$.
We also plotted the generalization error
obtained by perceptron learning
with learning rate $g=1$.
When $a=a_{\rm c1}$, the
generalization error under the perceptron algorithm
becomes equal to random guess $({\epsilon}_{\rm g}=0.5)$.
}
\Figure{
The optimal order parameter $R$ which gives the global minimum,
namely, the optimal generalization error ${\epsilon}_{\rm opt}$.
The system shows a discontinuous phase transition at
$a=a_{\rm c2}=0.80$ from the phase described by $R=1$
to the phase described by $R=R_{*}$.
We also plotted $R=1-2{\Delta}$
obtained by the perceptron learning with learning rate $g=1$.
When $a=a_{\rm c1}$, the overlap between the
teacher and student vanishes.
}
\Figure{
Flows of the order parameters $R$ and $l$
for the learnable case $(a=\infty)$ by the perceptron learning.
If one starts from large $l$,
the student begins to generalize after the length of the
weight vector $l$ decreases to some value.
}
\Figure{
Flows of the order parameters $R$ and $l$ for the unlearnable
cases $a=2.0$ by the perceptron learning.
The flows are attracted to a fixed point.
}
\Figure{
Flows of $R$ and $l$ for $a=\infty, 2.0$ and $0.5$
by the Hebbian learning.
For the cases of $a=\infty$ and $2.0$,
$R$ reaches $1$ and $l$ goes to $\infty$.
On the other hand, for $a=0.5$, $R$ reaches
$-1$ as $l$ goes to $\infty$.
}
\Figure{
Generalization error ${\epsilon}_{\rm g}$ for
$a=\infty, 2.0$ and $0.5$ by the Hebbian learning.
For $a=\infty$ and $2.0$,
the generalization error converges to
the optimal value $2H(a)$.
However, in the case of $a=0.5$,
the generalization error begins to
increase when the student learns too much (over-training).
}
\Figure{
Generalization error
for the unlearnable case $a=0.5$ with output noise
$\lambda=0.01$ and $0.20$ by the Hebbian learning. 
}
\Figure{
The trajectories in the $R$-$l$ plane with the
optimal learning rate by the perceptron learning
for $a=2.0$.
We choose the initial condition as
$(R_{\rm init},l_{\rm init})=
(0.01,0.10), (0.01, 1.00)$ and $(0.01, 2.00)$.
}
\Figure{
Same as figure 12 with $a=0.5$.
}
\Figure{
Generalization error for $a=2.0$
with the optimal learning rate $g_{\rm opt}$.
}
\Figure{
Same as figure 14 with $a=0.5$.
If we select a negative value as
the initial condition of $R$ for $a=0.5$,
the generalization error converges to $1-2H(a) (>0.5)$.
}
\Figure{
Generalization error of the Hebbian learning 
with queries for $a=1.0$. 
Over-training disappeared and the
generalization error converges 
to its optimal value. 
}
\Figure{
Generalization error of the Hebbian
learning with a weight-decay term for $a=0.5$.
Over-training disappeared and
generalization error converges to its optimal value.
}
\end{document}